\documentclass[10pt,onecolumn]{IEEEtran}
\usepackage{cite}
\usepackage{amsmath,amssymb,amsfonts}
\usepackage{algorithmic}
\usepackage{graphicx}
\usepackage{textcomp}
\usepackage{booktabs}
\usepackage{balance}
\usepackage{caption}
\usepackage{subfigure}

\begin{document}
%\history{Date of publication xxxx 00, 0000, date of current version xxxx 00, 0000.}
%\doi{10.1109/ACCESS.2017.DOI}

\title{A RISC-V SOC for Terahertz IoT Devices: Implementation and design challenges}
\author{\IEEEauthorblockN{\dag Xinchao Zhong,  \dag Sean Longyu Ma,  Hong-fu Chou\\}
\IEEEauthorblockA{\textit{Interdisciplinary Centre for Security, Reliability and Trust (SnT), University of Luxembourg, Luxembourg}\\ 
\textit{\dag School of Computer Science, The University of Auckland, New Zealand}\\ 
Corresponding author: Hong-fu Chou Email: hungpu.chou@uni.lu}

}
%\author{\uppercase{Xinchao Zhong}\authorrefmark{1}, \IEEEmembership{Member, IEEE},
%\uppercase{Sean Longyu Ma}\authorrefmark{1},
%\uppercase{Hong-fu Chou}\authorrefmark{2}, \IEEEmembership{Member, IEEE}, and \uppercase{Symeon Chatzinotas}\authorrefmark{2},
%\IEEEmembership{Fellow, IEEE}
%}
%\address[1]{School of Computer Science, The University of Auckland, Auckland 1142, New Zealand }
%\address[2]{Interdisciplinary Centre for Security, Reliability and Trust (SnT), University of Luxembourg, Luxembourg}
%\corresp{Corresponding author: Chiu-Wing Sham (e-mail: b.sham@auckland.ac.nz).}
%\tfootnote{This paragraph of the first footnote will contain support 
%information, including sponsor and financial support acknowledgment. For 
%example, ``This work was supported in part by the U.S. Department of 
%Commerce under Grant BS123456.''}

%\markboth
%{Xinchao \headeretal:A RISC-V SOC for Terahertz IoT Devices: Implementation and design challenges }
%{Xinchao \headeretal:A RISC-V SOC for Terahertz IoT Devices: Implementation and design challenges }

%\corresp{Corresponding author: First A. Author (e-mail: author@ boulder.nist.gov).}
\maketitle
\begin{abstract}
Terahertz (THz) communication is considered a viable approach to augmenting the communication capacity of prospective Internet-of-Things (IoT) resulting in enhanced spectral efficiency.  %To address the limitations of PWM-DACs, scholars have developed other DAC designs such as Pulse-count Modulation DACs (PCM-DACs) and First-order-noise Shaping DACs (FONS-DACs). Nevertheless, it is crucial to acknowledge that these alternate designs also have their own unique drawbacks. 
This study first provides an outline of the design challenges encountered in developing THz transceivers. This paper introduces advanced approaches and a unique methodology known as Modified Pulse-width Modulation (MPWM) to address the issues in the THz domain. In this situation involving a transceiver that handles complex modulation schemes, the presence of a mixed signal through a high-resolution digital-to-analog converter (DAC) in the transmitter greatly contributes to the limitation in maintaining linearity at high frequencies. The utilization of Pulse-width Modulation-based Digital-to-Analog Converters (PWM-DACs) has garnered significant attention among scholars due to its efficiency and affordability. However, the converters' performance is restricted by insufficient conversion speed and precision, especially in the context of high-resolution, high-order modulation schemes for THz wireless communications. The MPWM framework offers a multitude of adjustable options, rendering the final MPWM-DAC highly adaptable for a diverse array of application scenarios. Comparative performance assessments indicate that MPWM-DACs have enhanced conversion speed compared to standard PWM-DACs, and they also provide greater accuracy in comparison to Pulse-count Modulation DACs (PCM-DACs). The study presents a comprehensive examination of the core principles, spectrum characteristics, and evaluation metrics, as well as the development and experimental validation of the MPWM method. Furthermore, we present a RISC-V System-on-Chip (SoC) that incorporates an MPWM-DAC, offering a highly favorable resolution for THz IoT communications. 
\end{abstract}


%\titlepgskip=-15pt

%\maketitle

\section{Introduction}
\label{sec:introduction}
The exponential growth in the quantity of Internet of Things (IoT) devices will inevitably lead to a substantial surge in wireless network congestion. In \cite{Xu2021}, the reason for this is that the next wireless systems must offer increased system capacity along with exceptionally dependable and low-latency communication. Additionally, they must possess more flexible connectivity features to accommodate the evolving requirements of the IoT network. RISC-V is an emerging technology that is increasingly being used in low-power IoT applications. The architectural expansions of RISC-V and the commercialization of System-on-Chips (SOCs) employing this architecture have contributed to the stability of these extensions. Additionally, the lowered manufacturing costs and cheaper prices for end-consumers have facilitated the commencement of commercialization. A wide range of additional RISC-V cores \cite{Jang2021} have been produced using this instruction set architecture, and devices based on RISC-V are now being developed for specialized applications like the IoT, wearables, and embedded systems. The embedded IoT platform's application processor carries out the many functions of the IoT application, such as reading, storing, processing, and deciding when to send sensor data. The widespread use of IoT applications and their often strict power efficiency needs have led to the development of a wide range of application processors. It is conceivable to encounter IoT processors with power consumption in the nanowatt range that are specifically engineered for battery-less applications. Furthermore, cores that target power consumption in the milliwatt and microwatt ranges are also accessible. A method in \cite{Amor2022} for developing ultra-low power software-defined radios and an instruction-set extension for the open-source RISC-V ISA have been developed by the authors. With this extension, challenging arithmetic operations utilized in the physical-layer protocols of IoT communication systems are intended to be accelerated.

Mobile networks in the future are anticipated to integrate nomadic, dispersed base stations that employ unmanned aerial vehicles (UAVs). These networks will function as a supplementary element of wireless networks, linking a significant number of individuals globally as well as a multitude of stationary and portable cyber equipment situated in diverse areas. Simultaneously, the data transmission speed that mobile devices can handle is continuously rising due to the implementation of advanced networks like 5G and enhancements made to current infrastructures. In \cite{Krner2014TowardsTC}, there will be a significant increase in the volume of data being transmitted daily. This will lead to congestion in the communication between base stations and the main network through traditional backhaul connections.  It is anticipated that the sub-6 Gigahertz (GHz) and millimeter-wave (mmWave) frequencies may not have the capacity to facilitate communication for these consumers. Therefore, Terahertz (THz) communication (0.1–10 THz) \cite{Sarieddeen2019} has been considered a promising method to address the aforementioned issue because of its exceptionally large bandwidth. When building THz transceivers, it is crucial to take into account three primary performance aspects in \cite{Heydari2021}. Operating in the THz frequency range offers a wide and untapped spectrum of frequency bandwidth. However, increasing the amount of bandwidth assigned to each user poses several architectural challenges.
\begin{enumerate}
    \item The bandwidth rises when there is a significant difference between the two resonance frequencies. The assumption of a linear or uniform phase response is no longer genuine, resulting in phase distortion and the introduction of noise within the required frequency range. Utilizing high-order bandpass matching circuits is advisable when the fractional bandwidth exceeds 20\%.
    \item The adoption of THz carrier frequencies and the presence of inter-user interference (IUI) resulting from the implementation of IoT users would pose significant challenges owing to the high path loss characteristics. By leveraging multiplexing benefits from the propagation of distinct signal streams across several unique paths in different spatial and polarization domains, high diversity or rank order in THz channels can be used to boost channel capacity or multi-user service. 
    \item Transceivers operating at the THz frequency perform modulation and demodulation using digital signal processing techniques. Consequently, the digital realm may restrict the capacity of the analog baseband, mixed-signal, and RF chain to process modulated signals with a wide range of values and a high ratio of peak power to average power (PAPR). In addition, the effective application of higher-order modulation necessitates the use of data converters with better resolution, long dynamic range RF chains, and local oscillators with reduced phase noise, all while taking high sensitivity and linearity into account.
\end{enumerate}
%This paper presents a novel modulation method denoted as Modified PWM (MPWM), with which DAC can have variable pulse density and number of edges. An MPWM module with $n$-bit counter can have $(n-1)$ kinds of configurations of pulse density and number of edges. Therefore, DAC designer can have greater flexibility to adapt their design to specific applications to achieve better results. MPWM-DACs have better speed and accuracy than PWM-DACs and less edge error than PCM-DACs and FONS-DACs. The following chapters are organized as follows:

The ongoing challenge discussed in \cite{Heydari2021} is to provide energy-efficient resolutions for the DAC/ADC and DSP elements of integrated transmitter and receiver chipsets, which need to handle data speeds beyond 50 Gbps. The back-end and mixed-signal processing use a costly commercial Arbitrary Waveform Generator (AWG) and a real-time oscilloscope. Their objective is to produce potent sub-channelized modulated signals externally. Subsequently, these signals are used to provide power to the transmitting side, adjust for any discrepancies, convert the modulated radio frequency signal into its original form, and retrieve the unprocessed data stream on the receiving side. From a practical standpoint, integrating an external AWG and a real-time oscilloscope into a front-end is not a power-efficient approach. Expanding a transceiver from a single element to a multi-antenna configuration for ultra-high data-rate applications in the same region would greatly exacerbate the situation. To be more precise, transceivers that use higher-order modulations have a lower RF bandwidth for a given data rate. However, they need a much better resolution and a higher sampling rate for the DAC and ADC relative to the signal baud rate. Also, in practice, the sampling rate of a Nyquist-rate data converter is varied by five to six times the baud rate for a better bit error rate. In addition, as the modulation intricacy rises, the necessary resolution of the data converter also increases, which becomes progressively more difficult to achieve as the data rate increases. The ADC at the receiving end and the DAC at the transmit end are the main parts that determine the mixed signal. This signal has to meet strict requirements while high-speed, high-order modulation is being generated and processed. The contributions of this study are summarised as follows:
\begin{enumerate}
    \item We present a comprehensive overview that covers the prior methodologies and obstacles associated with the design of THz IoT devices, including the design of multi-antenna arrays and transceivers.
    \item By implementing a novel modulation technique called Modified PWM (MPWM), DACs are capable of displaying pulse density and edge count that can be adjusted. An MPWM module that incorporates an n-bit counter possesses the capacity to accommodate (n-1) distinct edge count and pulse density configurations.
    \item For the purpose of attaining exceptional outcomes, this approach provides DAC designers with enhanced versatility to modify their designs to suit specific applications.
    \item We present the design of an MPWM-DAC and RISC-V SoC specifically tailored for implementation in THz IoT applications.
\end{enumerate}
Section II presents a comprehensive examination of the design of high-performance THz antennas, both on-chip and off-chip, as well as multi-antenna arrays. In Section III, we tackle the task of developing multi-antenna transceivers for THz frequencies.
Section IV provides a detailed explanation and comparison of the concept and spectrum of MPWM with other modulation techniques. The performance of MPWM-DAC encompasses static error, integral nonlinearity, differential nonlinearity, and dynamic features. Section V presents the design and FPGA testing of MPWM-DACs, as well as the design of an MPWM-DAC and a RISC-V SoC for THz IoT applications.
\section{High-throughput THz multi-antenna array design}
Antennas play a crucial role in connecting the communication system to the surrounding environment.  On-chip antennas \cite{Aliba2022} offer a prospective option that is more advantageous than off-chip antennas for THz communication. An on-chip antenna can greatly simplify the matching network needed to connect the antenna to the RF circuitry. Simplifying the matching network can significantly enhance the system's performance by decreasing the loss and noise figure of the front end. In order to achieve optimal radiation or prolong the battery life of the systems, it is crucial to maximize the effectiveness of the antenna. In addition, the size of the on-chip antenna is the primary influence in defining the chip area. Therefore, it is imperative to minimize its dimensions to reduce manufacturing costs.
Antenna downsizing can be achieved by utilizing compact arrangements and incorporating materials with high permittivity.
\subsection{On-chip antennas}
\begin{table*}
\centering
\caption{Summary of the state-of-the-art THz antennas on chip.}
 \begin{tabular}{||c c c c c c  ||} 
 \hline
 Paper & \begin{tabular}[c]{@{}c@{}}Technology\end{tabular} & \begin{tabular}[c]{@{}c@{}}Frequency (GHz)\end{tabular} & Efficiency (\%) & \begin{tabular}[c]{@{}c@{}}Directivity (dBi)\\  \end{tabular} & \begin{tabular}[c]{@{}c@{}}Gain (dBi)
 \end{tabular}   \\ [0.5ex]  \hline\hline
 \small{\cite{Jalili2019}} & 0.13 $\mu$m SiGe &  370 & 40 & 8 & 4  \\
  \hline
 \small{\cite{Alib2019}} & 50 $\mu$m &  300 & 65 & NA & 4.5 \\ 
 \hline
  \small{\cite{Yi2021}} & 65-$nm$ CMOS &  260-280 & 15 & 7 & 22 \\ 
 \hline
   \small{\cite{Jalili2020}} & 65-$nm$ CMOS &  438–479 & 32 & 21.4 & 12.4   \\ 
 \hline
   \small{\cite{Li2017}} &  0.18 $\mu$m SiGe &  340 & 74  & NA & 7.9  \\ 
 \hline
    \small{\cite{Schmalz2013}} & 0.13 $\mu$m SiGe &  235-255 & 75 & NA & 7  \\ 
 \hline
    \small{\cite{Khan2015}} &  130 $nm$ SiGe BiCMOS & 165-175 & 45 & NA & 5  \\ 
 \hline  
    \small{\cite{Sarkas2012}} & 0.13 $\mu$m SiGe &  120 & 50 & NA & 6 \\ 

\hline
\end{tabular}
\label{table: related-works}
\vspace{-3mm}
\end{table*}
The rectangle, dipole, bow-tie, and slot antennas are commonly used topologies in on-chip design at frequencies above 300 GHz, with a gain above 2 dBi \cite{Tousi2015}. The patch antenna in \cite{Jalili2019}, created using 0.13 µm SiGe technology, operates at a frequency of 370 GHz. It achieves a gain of 4 dBi, an efficiency of 40\%, and a directivity of 8 dBi. The gain enhancement can be due to the increased electrical thickness of the dielectric as the frequency increases. 
Within the high gain zone, the return loss has an elevated value. Measuring the antenna above 100 GHz using on-probe methods gets difficult because the metallic elements of the probe station reflect the incoming radiated waves. This leads to a certain degree of discrepancy in the measured gain outcomes. In addition, the use of highly delicate probe tips during the measurements restricts the capacity to freely alter any components installed on the probe station that are essential for facilitating the measurement procedure, such as the mobility of the receiving reference antenna. 

The antenna outlined in \cite{Alib2019} consists of a square patch antenna positioned on a silicon substrate, accompanied by a ground plane. The substrate-integrated waveguide (SIW) is created by a patch that incorporates two T-shaped slots and short-circuited edges using metal vias. This method increases the size of the aperture region and reduces losses caused by surface waves and substrate. As a result, there is an enhancement in impedance matching, bandwidth, isolation, gain, and radiation efficiency. This arrangement minimizes the losses resulting from surface waves and the silicon dielectric substrate. The structure may be activated by employing two coaxial ports that are linked to the patch from the lower side of the silicon substrate. The enhanced aperture area significantly enhances both the impedance bandwidth and radiation characteristics within the frequency range of 0.28 THz to 0.3 THz. The antenna has a mean gain and efficiency of 4.5 dBi and 65\%, respectively. Furthermore, it possesses a self-contained configuration that exhibits exceptional isolation, surpassing 30 dB between the two ports. The on-chip antenna is 800 $\times$ 800 $\times$ 60 $\mu m^3$ in size. In addition, the antenna in \cite{Aliba20192} was constructed on a GaAs substrate that had a thickness of 0.5µm. The transceiver consists of a Voltage-Controlled Oscillator (VCO), buffer amplifier, modulator stage, power amplifier, frequency-tripler, and an on-chip antenna. The on-chip antenna utilizes substrate-integrated waveguide (SIW) technology and has a 4$\times$4 configuration of slots in both the longitudinal and transverse orientations, making use of metamaterial technology. The SIW antenna utilizes a high-pass filter to efficiently eliminate undesired harmonics and transmit the desired signal. The on-chip antenna has dimensions of 2 $\times$ 1 $\times$ 0.0006 $mm^3$ and demonstrates a minimum gain of 0.25 decibels isotropic (dBi), an average gain of 1.0 dBi, a minimum efficiency of 46.12\%, and an estimated efficiency of around 55\%. The transceiver emits an average output power of -15 dBm within the frequency range of 0.3-0.31 THz, which makes it highly suitable for near-field imaging applications.

The antenna presented in \cite{Yi2021} describes an antenna design that integrates an SIW cavity with a wideband dual-slot antenna on a chip. This design was specifically created for the use of frequency-modulated continuous-wave (FMCW) radars. Specifically, the existence of dual slots induces two resonance modes, while the SIW cavity induces supplementary modes that aid in achieving a broad bandwidth. The strength of this antenna varies from -1 dB to 0 dB across the frequency range of 260-280 GHz. It has an efficiency of 15\% for impedance matching, which corresponds to a fractional bandwidth. However, due to the substrate's thinness, electric fields are mostly limited to the area between the patch and the ground. As a result, both the gain and bandwidth are reduced. 

To create the radiation front end, one can utilize conventional antenna designs, such as rectangular patches and dipoles \cite{Maktoomi2023}.  Given that the substrate thickness usually falls within the range of 250-300 $\mu m$, it is possible to position the ground layer beneath the substrate to augment the total thickness between the antenna and the ground layer. The lens-integrated on-chip antennas \cite{Jalili2020}, which have been recently described, exhibit superior gain and bandwidth performance. The lens is responsible for improving the impedance matching with the air and also for collimating the output beam.  However, these methods necessitate the attachment of a large lens on their rear side. Antennas employing dielectric resonators (DR) \cite{Li2017} have a bandwidth reduction of less than 15\%, but make up for it with a higher gain of almost 6 dBi. By employing a dielectric material that is a minimum of 400 µm in thickness above the antenna, these on-chip DR antennas increase their surface area. 

According to the reference \cite{Schmalz2013}, using the localized backside etching (LBE) method on a group of folded dipole antennas made using 0.13 $\mu m$ SiGe technology results in a gain of more than 7 dBi and an efficiency of over 75\% within the frequency range of 235-255 GHz. Applying the LBE approach described in \cite{Khan2015} to eliminate the lossy silicon substrate around an on-chip antenna patch leads to a notable enhancement in performance. Nevertheless, the LBE method requires supplementary protocols and a designated area for etching, which must encompass the patch. According to \cite{Ng2018}, the substrate surrounding the patch is vulnerable to data loss. In order to address this concern, a Lossless Back End (LBE) method is implemented to eradicate the substrate responsible for data loss. On the other hand, this approach necessitates supplementary processing procedures and an extensive etching area.
This technique offers various methods for integrating lens antennas onto chips by achieving a sufficient balance in bandwidth through the inclusion of a sizable lens. Positioning the ground layer beneath the substrate enhances the electrical thickness of the dielectric. A considerable improvement in the amplification is observed. The use of a folded dipole antenna allows for achieving a maximum gain of 5 dBi, together with a radiation efficiency of 45\%, over the frequency range of 165-175 GHz. 

An improvement in radiation efficiency and a reduction in the space-wave quality factor occurs when the depth of the superstate is a multiple of a quarter-wavelength and an odd number. Thus, placing a high-dielectric superstrate over the antenna, as explained in \cite{Edwards2012}, is an alternate approach to enhance radiation efficiency. An elliptical slot antenna is manufactured using 0.13 µm CMOS technology. This antenna is placed on the top metal layer, while a ground layer is located at the LY metal layer, which is positioned 11 $\mu m$ below the top metal layer. The antenna is attached to a 400 $\mu m$  thick quartz superstrate. At 90 GHz, the antenna has a measured bandwidth of 3.9\%, a peak gain of 0.7 dBi, and a peak radiation efficiency of 30\%. However, the research carried out by \cite{Sarkas2012} shows that by adding a quartz superstrate and a parasitic patch to a shorted patch antenna on a 0.13 $\mu m$ SiGe substrate, a notable improvement of 6 dB may be achieved. The system operates at a frequency of 120 GHz and achieves an efficiency of around 50\%. Additionally, the system has a gain bandwidth of 9 GHz, with a 1-dB increase within this bandwidth. Enhancing the antenna's performance is achieved by utilizing a small superstrate, which increases the complexity of the manufacturing process. 

The unit cell of the artificial magnetic conductor (AMC) layer is constructed by enclosing pairs of walls, each composed of a perfect electric conductor and a perfect magnetic conductor. Through a waveport positioned above the unit cell, a planar wave strikes perpendicularly to the unit cell. It is possible to get an AMC layer surface reflection phase of zero close to the center frequency by carefully changing the sizes of the unit cell elements. Furthermore, to mitigate loss in the AMC layer, it is necessary to decrease the magnitude of the reflection coefficient. The bandwidth of a unit cell is defined as the range of the reflection phase, which varies between -90 and 90 degrees. Therefore, the design of a double-rhomboid bow-tie (DRBT) slot antenna with a back-to-back E-shaped slot and an anisotropic magnetic conductor (AMC) layer is implemented using a 0.13 $\mu m$  BiCMOS technology as presented in  \cite{Saad2019}. The antenna operates in the W-band frequency range of 75-110 GHz. The maximum gain observed is -0.58 dBi and the bandwidth is 6 GHz.
\subsection{off-chip antennas}
Integrating chipsets with off-chip antennas using normal interface techniques is limited by strict frequency limitations. Off-chip antennas are frequently used on Printed Circuit Boards (PCBs) when larger dielectric layers are necessary. The reduced manufacturing expenses associated with off-chip antennas enable the development of extensive off-chip antenna arrays, resulting in significantly amplified levels of radiated power. These antennas offer superior efficiency, gain, and bandwidth compared to on-chip antennas. While sacrificing bandwidth, the antenna described in \cite{Frank2019} utilizes numerous parallel leaky-wave structures to enhance the gain.
The antenna off-chip featured in \cite{Bae2021} achieves a high level of amplification and a broad range of frequencies by combining a multi-layered board technology with an L-probe feedline. 

It is crucial to have broadband antenna designs that can be accommodated within a limited number of metal layers and thin dielectric layers in order to manufacture antennas using commonly accessible rigid or flexible PCBs. The antenna presented in \cite{Maktoomi2023} reveals a stacked patch antenna that is connected to an external cavity backplane and operates in the frequency range of 91.5-134 GHz. The fractional bandwidth of the object is around 38\% and it has a peak gain of 8.1 dBi. This antenna demonstrates the greatest fractional bandwidth compared to off-chip antennas operating over 100 GHz. In addition, the ground and patch layers both utilize copper cladding with a thickness of 12 $\mu m$. The dielectric substrates have a relative permittivity ($\epsilon_r$) of 2.6 and a loss tangent of 0.003. In light of the aforementioned methodologies, the viewpoint about off-chip design problems can be succinctly summarized as follows:
\begin{enumerate}
    \item In order to provide broad frequency compatibility between the input G-S-G port and the slot, the feedline is constructed as a transmission line consisting of two sections. The 50 $\Omega$ impedance part is connected to the 50 $\Omega$ G-S-G port, while a larger low-impedance piece is connected to the impedance of the slot aperture. Undesirable parallel plate modes, characterized by resonance at different frequencies, arise when conductor layers are situated both above and below the feedline.
    \item In order to enhance the performance of wideband systems, the use of FPC technology permits a maximum thickness of 50 $\mu m$ for the substrate layers Sub1, Sub3, and Sub4. Sub2, with a reduced thickness of 25 µm, produces substantial coupling between the feedline and slot aperture. The G-S-G probe port is situated within the Gnd2 layer and is linked to the feedline. The feedline is a grounded co-planar waveguide designed to be compatible with the G-S-G probe port.
    \item The process of cavity generation entails the utilization of via arrays to construct a conductor enclosure surrounding patches. The manufacturing process can cause surface waves to pass through even the tiniest gaps between vias. Developing a broadband antenna is a significant challenge in this matter. At higher frequencies in a wideband antenna, the wavelength becomes shorter, necessitating a greater proximity between vias compared to lower frequencies.
\end{enumerate}

The primary issue associated with off-chip antennas is effectively guiding signals from the semiconductor to the antenna structure. The interface, equipped with pads on both ends, creates a network that exhibits self-inductance as well as capacitance and resistance. Ideally, this combination should operate as a direct path for electric current. The capacitors and inductors in this network resonate together at high frequencies, establishing a frequency limit called the self-resonance frequency (SRF). Regarded as the prevailing method for packaging in \cite{Simsek2020} and \cite{Simsek20202}, aluminum or gold wirebonds can also be employed to provide a connection between the chipset and antennas on the PCB. Standard wires typically have a diameter ranging from 10 to 75 $\mu m$. In high-power applications, many wirebonds may be used in tandem. In the usual wirebonding procedure, the wire is aligned at an almost right angle to the pads at one end. The wire becomes longer than required, leading to substantial self-inductance. The wirebonds exhibit a range of self-inductance values, which can span from a few hundred picohenries to a few nanohenries. Consequently, the wirebonds exhibit considerable reactance at higher frequencies, leading to a large reduction in their insertion loss and insufficient impedance matching.

\subsection{On-chip and off-chip antenna array}
As the frequency rises, the dielectric thickness increases as a percentage of the wavelength, while each antenna element's area decreases, increasing the gain bandwidth of the on-chip antennas. Following this logical sequence, this section explores off-chip and on-chip antenna arrays for multi-antenna systems in cases where employing a single-element radiator is impractical. In the THz frequency band, a chip with a single antenna element generates maximum power that covers a limited distance. Antenna arrays are utilized to enhance the coverage area of a communication system by greatly amplifying the total radiated power and enhancing the beam's directionality. The antenna components are spaced at a distance equivalent to half the wavelength, enabling a coherent combination of radiated waves and limiting interference.

Using an on-chip antenna array simplifies routing and matching between transceiver circuitry and antenna components. As previously indicated, on-chip antenna performance improves with higher frequencies, with most arrays operating beyond 150 GHz. The authors in \cite{Han2013} implement an 8-element slot array with a 10 $mm$ Si lens at the bottom in a 65 $nm$ CMOS process. Simulated directivity is 16.6 dB, the overall radiation efficiency is 42\%, and the impedance bandwidth exceeds 60 GHz in the 260 GHz range. The eight-element on-chip antenna array in \cite{Yang2016} features a $\lambda$/4 thick Quartz superstrate on top. The Quartz superstrate boosts gain by 3.1 dB and efficiency from 22\% to 45\% for each antenna element. This leads to an array gain of 11-12 dBi and a 10 GHz bandwidth at 385 GHz frequency. At 412-416 GHz, a 4 $\times$ 4 patch antenna in an oscillator network obtains a peak effective isotropic radiated power (EIRP) of 14 dBm in \cite{Saeidi2020}.

Enhanced gain and bandwidth performance can be achieved by scaling antenna size and accessing thicker substrates in off-chip technologies. The allocation of space for an antenna on a PCB board is not a major issue, in contrast to an on-chip antenna, where it occupies a substantial portion of the chip. This enables the use of larger off-chip antenna arrays, hence improving antenna gain performance and transceiver chip EIRP. The diminished amplification and frequency range capabilities shown in the following investigations can likely be attributed to the reduced thickness of the interposer layer and the scattered connection between the chipset and the interposer. The designs in \cite{Herrero2009} are pioneering examples of off-chip antenna arrays operating at frequencies above 100 GHz. This work constructs two patch arrays on a RO3003 substrate with a dielectric constant ($\epsilon_r$) of 3. One array is fed in series, while the other is supplied corporately. The serial-fed antenna has a maximum gain of 6 dBi and a frequency range of 5 GHz centered at 122 GHz. On the other hand, the corporate-fed antenna has a gain of 5 dBi and a bandwidth of 7 GHz. 

To showcase long-distance communication capabilities, the authors in \cite{shahramian2018} proposed the use of a 384-element array operating at a frequency of 90 GHz. This array would consist of 11 layers of metal stacks on a PCB, and would be linked to the chipset by flip-chip interconnects. The array achieved an impressive EIRP of 60 dBm within a conservative 20 GHz bandwidth. The authors in \cite{Naviasky2021} constructed 16 patch antennas on an interposer board positioned between the chipset and PCB to facilitate large-scale MIMO applications. This study employed an interposer board to establish a connection between narrow signal lines on the chipset and wider lines on the board. The bandwidth for return loss was measured to be between 71 and 84 GHz, while the peak gain was about 5 dBi. 

The on-chip antenna array in \cite{Alibak20193} operates within the frequency range of 0.450-0.475 THz. It utilizes two vertically oriented DRs on a silicon substrate, employing standard CMOS technology. To reduce energy dissipation, one can create a winding pathway in the silicon substrate and enclose it with a metallic barrier, thereby reducing substrate loss and surface wave effects. The integration of slots and vias results in the antenna adopting a metamaterial structure that occupies a very small area. The dimensions of the antenna are 400 x 400 × 135 $\mu m^3$. The antenna achieves a peak gain of 4.5 decibels relative to an isotropic radiator and has a radiation efficiency of 45.7\% at a frequency of 0.4625 THz. A series-fed double-DR on-chip antenna array is a promising candidate for THz integrated circuits. By employing CMOS 20µm silicon technology, the authors in \cite{Aliba20194} develop an innovative on-chip antenna array designed for operation at frequencies ranging from 0.6 to 0.65 THz. This array configuration has three vertically aligned layers of Silicon-metal-Silicon. The intermediate metal layer functions as a ground plane sandwiched between two silicon layers. The uppermost layer consists of a pair of antennas, each equipped with three interconnected radiating elements. Radiation elements exhibit the behavior of linked dual rings, like a metamaterial. This arrangement increases the effective aperture area of the array. The inclusion of metallic via-holes between radiation elements in three layers mitigates the effects of surface waves and substrate losses. The antenna is operated via microstrip wires that are open-circuited on the rear of the structure. The ground-plane layer incorporates slots that facilitate the transmission of electromagnetic energy from the lower layer to the radiating components located on the upper layer. The dimensions of the antenna array are 0.4$\times$ 0.4 $\times$ 0.06 $mm^3$. The on-chip antenna array in \cite{Aliba20194} attains an average radiation gain, efficiency, and isolation of 7.62 dBi, 32.67 \%, and negative 30 dB, respectively. The results confirm the effectiveness of the antenna array for THz-integrated circuits.

In essence, on-chip antennas that operate at frequencies over 150 GHz adhere to radiation laws, enabling the creation of a fully integrated communication system.  On-chip antennas are frequently used at lower THz frequencies due to their ability to provide a broad gain bandwidth while minimizing substrate elevation. Although there have been improvements in radiation quality, the linkage between the chipset and off-chip antenna remains a major obstacle to attaining optimal off-chip radiation frequencies. Wirebonds have traditionally been employed to establish connections with frequencies above 100 GHz. Furthermore, flip-chip technologies \cite{beica2013flip}, namely copper-pillar and solder bump, offer an impressive SRF of around 200 GHz. This makes them highly suitable for the development of future THz communication systems.
\section{THz multi-antenna transceivers Design}
The majority of studies in multi-antenna transceiver design have focused on beamforming using phased-array solutions. Following the first showcases of single, four, and eight-element arrays on a single microchip, the phased-arrays rapidly progressed to 16 and 32-element configurations on a single microchip. In the transmit mode of a phased array with $N$-elements presented by the following principle in \cite{Shin2013}, the effective isotropic radiated power (EIRP) is defined as the product of the total transmit power $P_t$ and the transmit antenna gains $G_t$. $P_t$ represents the power emitted per element, while Gt is directly proportional to $N$. The EIRP, on the other hand, is directly proportional to $N^2$. When operating in receive mode, the gain of the phased array antenna is directly proportional to the number of elements, denoted as $N$. Hence, the link budget, which is directly proportional to the product of $P_t$, $G_t$, and receiver antenna gain $G_r$, exhibits a cubic dependency on the number of $N^3$. This section will discuss two designs: high-performance multi-antenna systems and low-complexity MIMO precoding designs. These architectures address the obstacles and concerns related to increasing bandwidth, modulation order, and transmit power in THz technology.
\subsection{High-performance multi-antenna systems}
Due to the reduced dimensions of passive components in the THz frequency range, it becomes feasible to consider the implementation of interconnected multi-antenna transceiver arrays. All-digital beamforming enables the transmission of several beams simultaneously, providing maximum flexibility and data transfer rate. The digital beamforming (DBF) technique has three main benefits. By employing digital precoding, it becomes possible to get precise resolution in both magnitude and phase. Moreover, a DBF array possesses the potential to increase its capacity by superimposing several beams to handle various data streams. In addition, the complete DBF architecture enables independent beamforming precoding on each subcarrier or resource block for multicarrier transmissions. This results in exceptional performance across a broad spectrum of frequencies. As the authors presented in \cite{Yang2018}, DBF-based millimeter-wave MIMO systems enable multi-user access and exhibit exceptional spectrum utilization. The principal constraints associated with the development of a DBF-based millimeter-wave MIMO transceiver are hardware complexity, financial investment, and power consumption. In order to facilitate the further development of millimeter-wave MIMO systems utilizing DBF, cost reduction in digital baseband processing will be the primary objective. Novel semiconductor manufacturing processes and improved integration techniques are assisting in the resolution of these constraints.

Lately, there has been a concentration on a hybrid beamforming approach that merges analog beamforming and digital MIMO coding. This technology offers a significant advantage by minimizing the complexity of the digital baseband through the utilization of a reduced number of up/down conversion chains in systems that have a high number of antennas. Consequently, it has emerged as a feasible choice for both outdoor and interior millimeter-wave/terahertz communication. By combining multi-beam digital baseband processing with analog beamforming, it is possible to simultaneously improve both multiplexing and beamforming gain. RF phase shifters are primarily used to alter the direction of the main lobe. RF variable gain attenuators and/or amplifiers (VGAs) provide interference spatial filtering by aligning the zero locations of each beamforming line with the incident angle of interference. The number of parallel data streams $K$ determines the minimum required number of RF chains $N_{RF}$ in a hybrid design. On the other hand, the beamforming gain is achieved by using $N_{RF}$ complex weighting factors that appear at each antenna. In the context of sub-array systems, hybrid beamforming have the capability to either receive or transmit an entire set of data streams from $N$ antennas when $N = N_{RF}$ or only a subset of data streams when partially interconnect RF circuits supply subarrays, i.e., $N_{RF} < N$, per antenna. A comprehensive array fulfills the function of an exclusively DBF, upon closer inspection. $N_{RF} \times N $ denotes the quantity of signal processing paths for the subarray, specifically from the digital baseband to the antenna front-end. $N^2$ signifies the total number of signal processing paths for the array. On the contrary, the beamforming gain of the sub-array is equivalent to the full array's $N_{RF}/N$. Consequently, in hybrid beamforming, the objectives of beamforming gain and signal processing complexity are in direct opposition. As the authors presented in \cite{Mondal2018}, it allows for the receiving of two streams by utilizing the Cartesian combining principle. The execution of this design for a two-stream reception necessitates the use of eight splitters, twenty combiners, and twelve mixers. The presence of several signal routes leads to electromagnetic cross-talk, which arises from the frequent cross-overs that take place between these paths.

Phased arrays provide a range of capabilities found in a multi-antenna system, such as enhancing capacity and diversity. As the authors presented in \cite{Natarajan2006}, the initial integration of the phased-array system in silicon involved the incorporation of local oscillator (LO) phase shifting. The SiGe transceiver has four components for both transmitting and receiving signals, as well as circuitry for generating and distributing LO frequencies. Additionally, it features a local design that allows for phase-shifting in the LO path, enabling beam steering. The main benefit of this architecture is that the phase-shifters are positioned at a distance from the RF route. Consequently, the implementation of the LO-path phase-shifting scheme at a local level allows for the creation of a reliable distribution network that can effectively handle THz frequencies and/or a larger number of components. The presence of large array sizes can present a notable design obstacle for the LO distribution network. Nevertheless, this worry can be mitigated by employing a phased-array transceiver that incorporates RF phase shifting. In \cite{Sadhu2017}, the described approach involves circuitry that allows for accurate manipulation of beam direction, phase, and amplitude at each individual front end. Additionally, independent control of tapering and beam steering is achieved at the array level. The integrated circuit is designed using 130-$nm$ SiGe BiCMOS technology. It consists of 32 transceiver components and supports simultaneous independent beams in two polarizations for both transmission and reception operations.

\subsection{Low-complexity MIMO precoding Designs}
Without investigating innovative architectural approaches, it may be unfeasible to include a unified ultra-high data-rate wireless transceiver that surpasses 50 Gbps. This is attributed to the substantial power consumption of the baseband units and data converter, which may reach a maximum of 10 watts. Utilizing a multi-antenna arrangement, rather than a single-element system will result in an increase in the power consumption of the transceiver beyond the frequency of 100 GHz. Streamlining the complexity of data converters and back-end digital signal processing (DSP) is essential, as it allows for the creation of energy-efficient and cost-effective high-speed wireless connections that can be accessed by a large number of customers. Conventional MIMO systems are solely operational within the baseband domain. However, full-digital precoders are impracticable due to the substantial energy consumption of high-frequency mixed-signal components and the attendant manufacturing costs. The primary reason for this is the considerable number of RF chains that are necessary for THz MIMO systems, which frequently comprise several tens to hundreds of antennas. An array of signals is processed using a hybrid cascaded RF precoder and baseband precoder. The RF precoder employs analog phase shifters to precisely manipulate the phase of signals entering and departing the antenna components. This enables the creation of multiple beams that align with the dominant THz channel paths. The baseband precoder exhibits more adaptability compared to the constant-gain/phase-only functionalities of the RF precoder. The substantial size of the antenna in the hybrid precoder architecture poses challenges in obtaining an optimum full-digital precoder. To develop a hybrid precoder, an optimal full-digital precoder must be employed. The complete digital precoder is often derived from the dominant singular vectors of a channel matrix in the explicit spatial domain. The computation of the singular value decomposition (SVD) for the explicit channel matrix is complex because of the abundance of antennas. In addition, receiver implementation of hybrid precoder designs is commonplace to reduce the quantity of feedback demanded. The execution of SVD ensues as a result. Determining the optimal full-digital precoder is the initial step in formulating the hybrid precoder design as a sparse optimization problem. One potential solution for the hybrid precoder design is to employ simultaneous orthogonal matching pursuit (SOMP)\cite{Cai2011}, in which the approach aims to select a suitable combination of analog beamforming vectors from a pre-determined set of options. Implementing this approach results in near-optimal performance. However, the hybrid precoder using the SOMP technique necessitates matrix inversion, resulting in a significant increase in complexity. It then computes the matching baseband precoding matrix in order to minimize the Euclidean distance between the current outcome and the ideal precoding matrix. The authors of \cite{Yu2016} proposed an alternating minimization (Alt-Min) approach to separate the hybrid precoder design into two distinct sub-problems. Initially, a temporary digital precoder is derived using the least square solution, while the initial analog beamformer is generated with random phases. Following this, the associated analog beamformer is modified via the phase extraction approach. The process of alternate iterations persists until the criterion stated by the user is met. The SOMP-based hybrid precoding can be interpreted as a method for sparse reconstruction, whilst the Alt-Min algorithm can be seen as a strategy for manifold optimization. 

Although there have been several suggestions for adopting hybrid precoding techniques, the VLSI implementation of precoding algorithms has not received substantial attention. The authors of the study described in \cite{Lee2015} established a technique known as parallel-index-selection matrix-inverse-bypass simultaneous orthogonal matching pursuit (PIS-MIB-SOMP). This approach seeks to avoid the necessity of performing a complete matrix inversion in the traditional SOMP algorithm. The researchers in\cite{Ho2019} developed a modified version of the SOMP technique called orthogonality-based matching pursuit (OBMP)\cite{Hung2015}. OBMP employs a discrete Fourier transform (DFT) codebook to replace the original candidate array solutions, hence reducing computing expenses. The authors of \cite{Chen2019} introduced an improved iteration of the orthogonal matching pursuit (OMP) method in their paper. The enhanced OMP algorithm integrated a revolutionary least-squares approach that employed QR decomposition. The authors also investigated the possible advantages of utilizing the Coordinate Rotation Digital Computer (CORDIC) method for implementing this algorithm within the framework of Very Large Scale Integration (VLSI). However, for these designs to serve as a reference design, an ideal precoding matrix is required. This matrix can only be constructed by using the SVD of the explicit channel matrix. However, this process significantly increases the complexity of the transceivers, making it impractically high. 

Inspired by the approach in \cite{Brady2013} to providing extensive bandwidths, short wavelengths at mm-wave or THz generate a spatial signal space with a high number of dimensions. This presents an opportunity to leverage high-dimensional MIMO techniques in order to achieve substantial capacity increases. Through the integration of a hybrid analog-digital transceiver and the beamspace MIMO communication concept, continuous aperture-phased MIMO attains performance that is nearly optimal while significantly reducing complexity. The researchers demonstrated in \cite{Chen2017} that beamspace singular value decomposition (SVD) is the most efficient approach for obtaining a full-digital precoder with minimal effort. It efficiently reduces power consumption for both the base station and user equipment. The authors utilize compressed sensing (CS)-based channel estimators to get reduced-dimension beamspace channel state information (CSI). This algorithm performs SVD implicitly on the reduced-dimensional beamspace channel instead of explicitly on the large-dimensional spatial domain channel. The total complexity is proportional to the number of antennas in the MIMO system, which is considerably higher than the channel sparsity. The CS-BHP technique described in \cite{Chen2017} achieves a considerable reduction in complexity compared to the state-of-the-art approach by 99.6\% by utilizing low-dimensional beamspace CSI. This eliminates the need for matrix inversion calculations and matching pursuit rounds, making it more efficient than an ideal full-digital precoder. Furthermore, it has a performance decrease of less than 5\%. The suggested design in \cite{Emre2021} utilizes a two-stage precoding approach. The first stage involves using a sparse matrix for precoding in the beamspace domain, followed by converting the outcome to the antenna domain using an inverse fast Fourier transform. This move is performed to simplify the complexity of multi-user (MU) precoding in all-digital base station systems. This approach uses OMP to calculate sparse precoding matrices in the beamspace domain. As a result, this technique reduces the complexity of precoding compared to standard linear antenna-domain precoders that involve dense matrix-vector multiplication. The output is preprocessed and then transformed into the antenna domain using an inverse fast Fourier transform (IFFT). The authors demonstrate that their methods provide a bit error-rate (BER) performance that is similar to that of traditional antenna-domain Wiener filter (WF) precoding, but with almost double the complexity. 

\section{High-speed high-resolution DAC for high order modulation}

%A Pulse-width Modulation-based Digital-to-Analog Converters (PWM-DAC) is fundamentally composed of a PWM generator and a filter, as outlined in previous studies \cite{hoeschele1994analog}\cite{staff1986analog}. The PWM generator is typically digital and can be an integrated module within a Microcontroller Unit (MCU) or a Digital Signal Processor (DSP). The filter is often a straightforward first-order RC filter. This simplicity in architecture contributes to the low cost and ease of implementation, which are the primary advantages of PWM-DACs. These attributes have garnered considerable attention in both academic and industrial circles. Numerous semiconductor companies specializing in MCUs and/or DSPs have released technical documentation on the design of PWM-DACs \cite{IXYSAN035701,zhou2010using, TIDSPRA490}.
As described in prior research, a Pulse-width Modulation-based Digital-to-Analog Converter (PWM-DAC) consists primarily of a filter and a PWM generator \cite{hoeschele1994analog}\cite{staff1986analog}. Integrated modules for Microcontroller Units (MCUs) or Digital Signal Processors (DSPs) constitute the PWM generator, which is principally digital. Frequently, the filter is a simple first-order RC filter. The primary benefits of PWM-DACs are their low cost and straightforward implementation, which are outcomes of their straightforward architecture. These characteristics have attracted significant interest in both academic and industrial spheres. A considerable number of semiconductor firms that specialize in DSPs and/or MCUs have published technical documents detailing the construction of PWM-DACs \cite{IXYSAN035701,zhou2010using, TIDSPRA490}.

%The performance of these DACs depends on the modulation method and filter design. Their resolution is limited by the bit width of the counter inside the modulator and the low pass filter’s performance. To design a DAC with $n$-bit resolution, the counter’s bit width must be more significant than $n$. Furthermore, the filter must have low cutoff frequency and/or high attenuation outside the pass band, which limits the speed and accuracy of these DACs
DAC performance is contingent on filter design and modulation technique. The bit breadth of the counter within the modulator and the efficacy of the low pass filter impose constraints on their resolution. In order to implement a DAC with a resolution of $n$ bits, the bit width of the counter must be greater than $n$. In addition, beyond the pass band, the filter must have a low cutoff frequency and/or high attenuation, which restricts the speed and precision of these DACs.

%Second-order low-pass Butterworth filters are the cost-effective choice for PWM-DAC, PCM-DAC, and FONS-DAC. In this case, for the worst ripple of $R_WC$ - in the unit of least significant bit (LSB), the approximate normalized cutoff frequency of the filter for PWM is \cite{Halper1996}.
Second-order low-pass Butterworth filters are the most economically viable option when it comes to PWM-DAC, PCM-DAC, and FONS-DAC. The normalized cutoff frequency of the PWM filter, expressed in the least significant bits (LSB), is approximate \cite{Halper1996} for the worst ripple of $R_WC$ in this instance.
\begin{equation}\label{cutoff_freq}
f_cT=0.9^2\sqrt{R_{WC}/2^n}.
\end{equation}
where $f_c$ is the cutoff frequency and $T=2^n/f_clk$ is the PWM period (n is the bit width of the counter inside the PWM and PCM circuit, and $f_clk$ is the working clock). 
%For a clock rate of 100 MHz, to implement a $12$-bit PWM-DAC with second order low-pass Butterworth filter, the cutoff frequency of the filter is 250 Hz and the settling time of the DAC is 0.76 ms, which means the conversion rate is under 1 kHz considering the settling time can only be a minor fraction of the conversion period. Furthermore, $f_cT$ decreased rapidly as the DAC’s resolution $ n$ increased, hence it will be more challenging to get a high-resolution PWM-DAC with reasonable speed.
The cutoff frequency of the second-order low-pass Butterworth filter for a $12$-bit PWM-DAC with a 100 MHz clock rate is 250 Hz, and the DAC's settling time is 0.76 ms; therefore, the conversion rate is less than 1 kHz, given that the settling time can only represent a negligible portion of the conversion period. Moreover, as the resolution (n) of the DAC increased, there was a significant decrease in $f_cT$; consequently, achieving a high-resolution PWM-DAC operating at a reasonable speed will be more difficult.

%PCM-DAC and first-order-noise shaping DAC (FONS-DAC) ease the stringent requirement for $f_cT$ by applying highly frequent pulses to raise the speed. However, PCM has a great number of rising and falling edges in a PCM period, which results in great static error due to the nonideal characteristics of the edges.
In order to increase the speed, PCM-DAC and first-order noise shaping DAC (FONS-DAC) reduce the stringent requirement for $f_cT$ through the application of extremely frequent pulses. Nevertheless, PCM periods contain a large number of ascending and descending edges, which leads to significant static error as a consequence of the edges' nonideal properties.

%For a $12$-bit PCM-DAC, the maximum error caused by the edge is $2^{11}\times\Delta W\times f_{clk}$, where $\Delta W$ is the pulse width deviation caused by the delay difference between rising and falling edge. Therefore, the pulse width deviation can be enlarged for 2048 times, which can reduce the DAC’s accuracy severely. In addition, with higher pulse density, PCM-DAC and FONS-DAC have the worst error performance.
The edge induces a maximum error of $2^{11}\times\Delta W\times f_{clk}$ for a $12$-bit PCM-DAC, where $\Delta W$ represents the pulse width deviation resulting from the delay differential between the rising and falling edges. Consequently, a 2048-fold increase in the pulse width deviation can significantly compromise the accuracy of the DAC. Additionally, PCM-DAC and FONS-DAC exhibit the worst error performance as pulse density increases.

\subsection{Principle and spectrum of MPWM}
\begin{figure}[htbp]
\centerline{\includegraphics[width=0.5\textwidth]{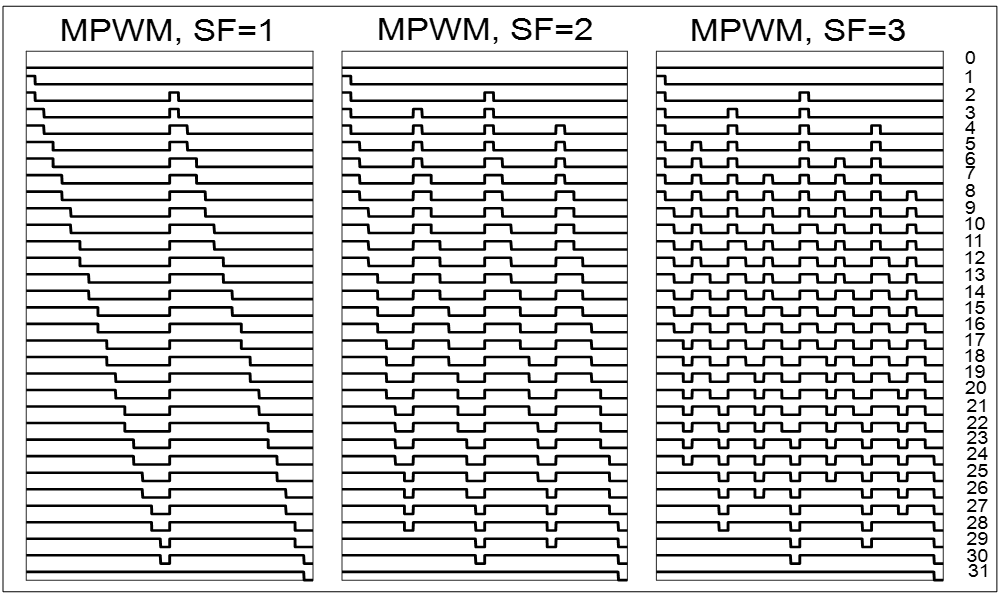}}
\caption{The waveform of unfiltered MPWM digital output.}
\label{wave_MPWM1}
\end{figure}

\begin{figure}[htbp]
\centerline{\includegraphics[width=0.5\textwidth]{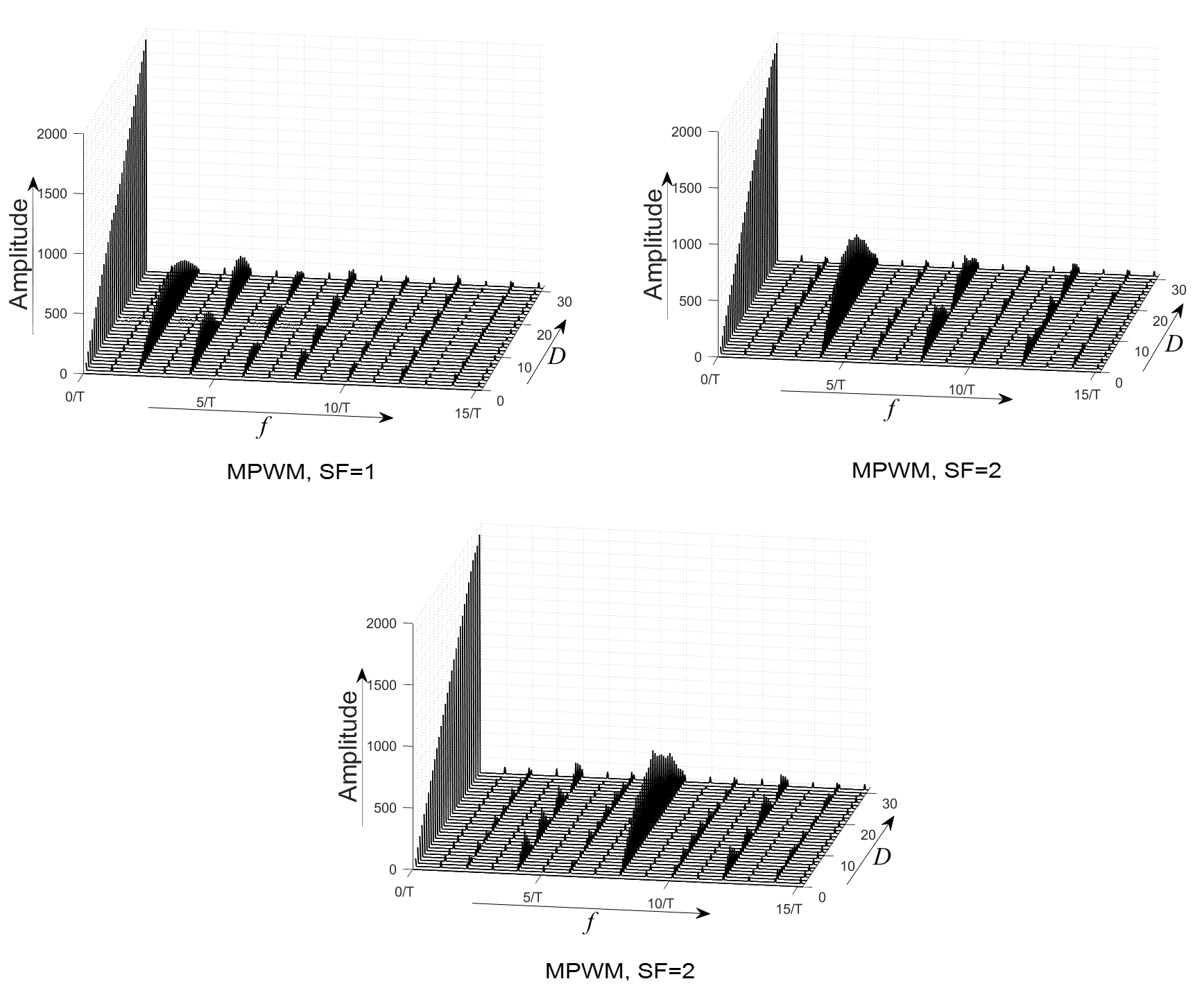}}
\caption{The spectrum of unfiltered MPWM digital output.}
\label{spectrum_MPWM1}
\end{figure}

%For a n-bit counter, the period of the MPWM signal is $T=2^n/f_{clk}$, where $f_{clk}$ is clock frequency. We introduce the splitting factor SF and the splitting number SN, which have the relationship of $SN=2^{SF}$. MPWM is split into SN sub-regions. For the $32$ clock period, the waveform of MPWM in the case of $n = 5$ and $SF = 1, 2, 4$ is shown in Fig. \ref{wave_MPWM1} and the corresponding spectrum in Fig. \ref{spectrum_MPWM1}.
The period of the MPWM signal for an n-bit counter may be calculated using the formula $T=2^n/f_{clk}$, where $f_{clk}$ represents the clock frequency. We define two terms: the splitting factor (SF) and the splitting number (SN). These terms are related by  $SN=2^{SF}$.
MPWM is divided into SN sub-regions. The waveform of MPWM for $n = 5$ and $SF = 1, 2, 4$ is displayed in Fig. \ref{wave_MPWM1} for a duration of 32 clock periods. The matching spectrum can be seen in Fig. \ref{spectrum_MPWM1}.
\begin{figure}[htbp]
\centerline{\includegraphics[width=0.5\textwidth]{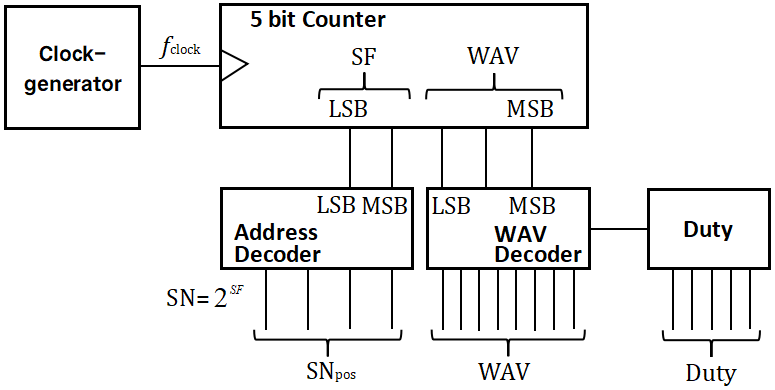}}
\caption{The MPWM generation using a 5-bit counter with 2 address bits.}
\label{5b2a_MPWM}
\end{figure}

\begin{figure}[htbp]
\centerline{\includegraphics[width=0.5\textwidth]{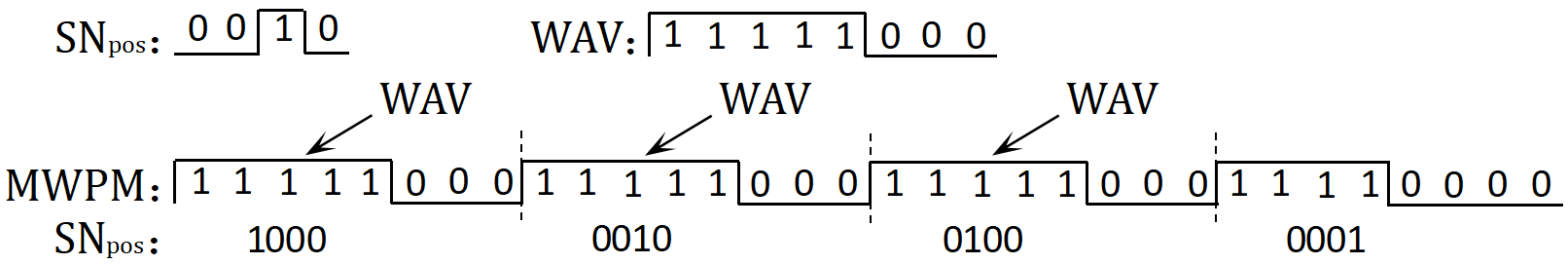}}
\caption{MPWM waveform in the case of Duty = 19, $SN_{pos}$ = 2, and Data = 4.}
\label{wav_example}
\end{figure}

%Fig. \ref{5b2a_MPWM} shows the generation principle of the MPWM wave. The lower SF bits of the counter represent the address (the sub-region in an MPWM period $T$), and the higher $(5-SF)$ bits represent the “wave” to be put in the sub-region. The address bits of the counter are output to a decoder (Address Decoder) with $SN$ bits output. When the output of the address decoder is a signal with a length of $SN$, which has a bit $1$ at position $SN_{pos+1}$. This signal indicates the position of the “wave” in the entire MPWM signal. 
Fig. \ref{5b2a_MPWM} illustrates the principle of generating the MPWM wave. The lower significant bits (SF) of the counter correspond to the address, which represents the sub-region inside an MPWM period $T$. The upper bits, namely $(5-SF)$ bits, indicate the specific "wave" that should be placed in the sub-region. The counter's address bits are sent to an Address Decoder, which produces an output of $SN$ bits. When the output of the address decoder is a signal of length $SN$ with a binary value of $1$ at location $SN_{pos+1}$. This signal represents the location of the "wave" inside the full MPWM signal. 
The higher $(5-SF)$ bits of the counter are connected to the “WAV decoder”, whose output WAV is a signal with a length of $2^{5-SF}$. \begin{figure}[htbp]
\centerline{\includegraphics[width=0.5\textwidth]{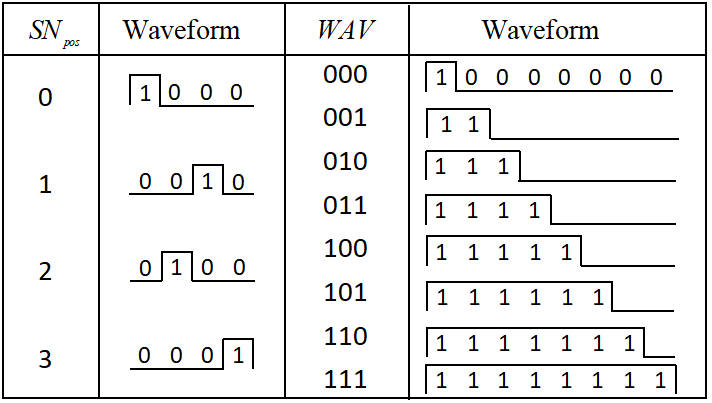}}
\caption{The waveform of the output of Address Decoder and Data Decoder.}
\label{wav_table}
\end{figure}
%Fig. \ref{wav_table} shows the waveform corresponding to WAV. The WAV Decoder is controlled by the Duty input. In an MPWM period, MPWM generates waveform according to Fig. \ref{wav_example} for “Duty” clock cycles, then it outputs bit $0$ until the next MPWM period begins.
Fig. \ref{wav_table} displays the waveform that corresponds to the WAV format.
The operation of the WAV Decoder is governed by the Duty input. During an MPWM period, the MPWM creates a waveform as shown in Figure \ref{wav_example} for a specified number of clock cycles called the "Duty". After that, it outputs a bit with a value of $0$ until the next MPWM period starts.
In order to examine the spectrum of MPWM, we will configure the counter to have a bit width of $5$ and establish a unit signal as follows:
\begin{equation}\label{x_unit}
x_{unit}= \left\{\begin{matrix}
1, ~~mT/32\leq t\leq (m+1)T/32
\\ 
0,~~~~~~~~~~else.~~~~~~~~~~~~~~~~~~~~~~~~~~~~~~
\end{matrix}\right.    
\end{equation}
which represents the signal at $t = mT /32$, the duration is $T /32$, and the period is T. By changing the value of m, $x_{unit} (t)$ can be used to form various MPWM waveforms. According to the Fourier series formula of the periodic signal as follows:
\begin{equation}\label{x_t}
x(t)=\sum_{k=-\infty}^{\infty}a_ke^{-jk\frac{2\pi}{T}t}dt.     
\end{equation}
\begin{equation}
a_k=\frac{1}{T}\int_{T}x(t)e^{-jk\frac{2\pi}{T}t}dt.     
\end{equation}
the Fourier series of $x_{unit} (t)$ is calculated as
\begin{equation}\label{a_k_fourier}
a_k=\left\{\begin{matrix}
1/32,~~~~~~~~~~~~~~~~~~~~~~~~~~~~~~~~~~~~~~k=0
\\ 
e^{-jk(\pi/32+m\pi/16)}sin(k\pi/32)/k\pi ,~~~k=1,...,31.
\end{matrix}\right.    
\end{equation}
\begin{figure}[htbp]
\centerline{\includegraphics[width=0.5\textwidth]{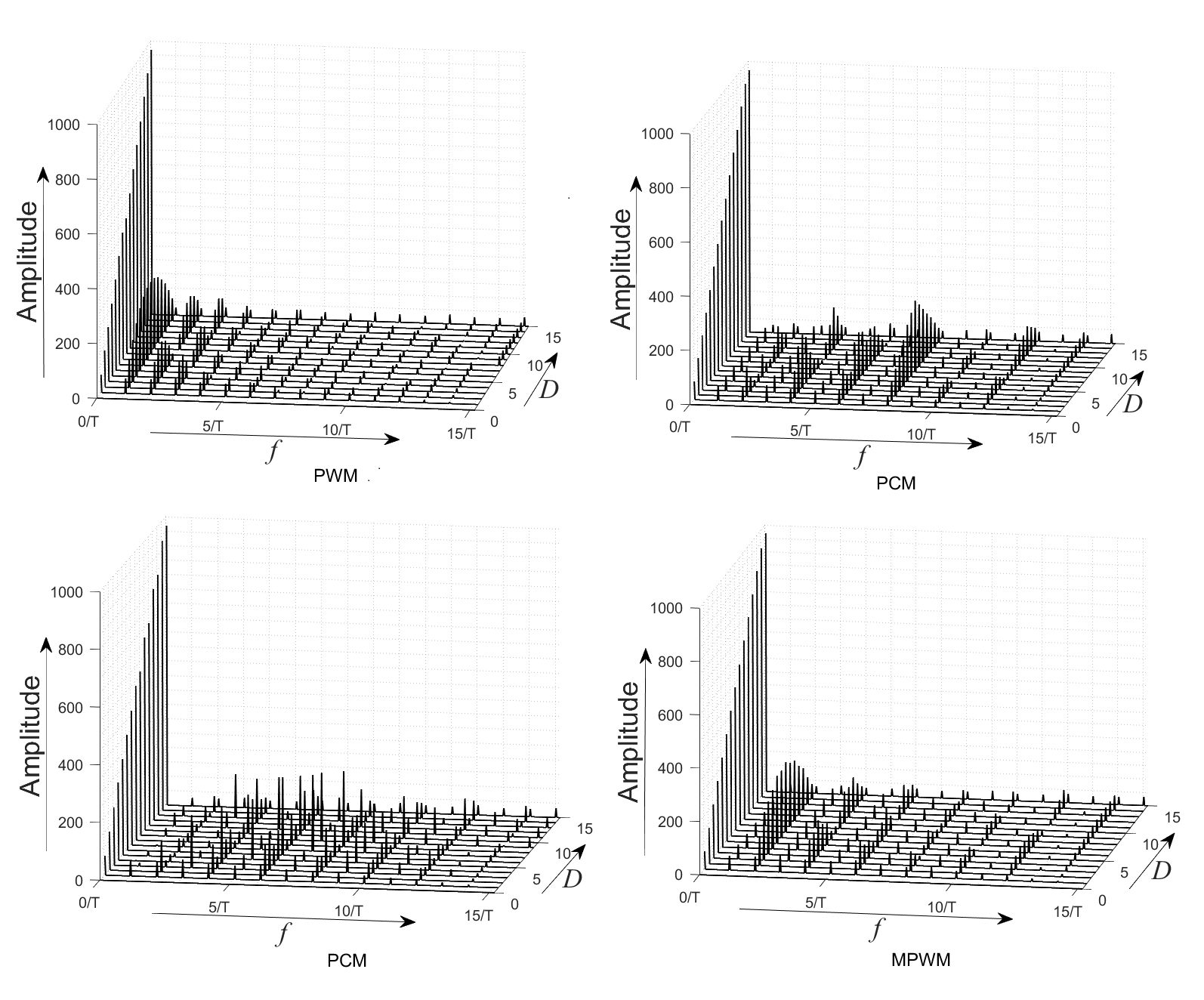}}
\caption{The Spectrum comparison of the unfiltered digital output.}
\label{spectrum_compare}
\end{figure}
By adding $x_{unit} (t)$ with different values of m, we can get MPWM with the splitting number $SN$ = 2, 4, and 8, respectively. According to the linear properties of the Fourier series, it is easy to calculate the Fourier series of these MPWMs. The frequency spectrum of PWM, PCM, FONS, and MPWM are shown in Fig. \ref{spectrum_compare}. The harmonic energy of the MPWM is mainly at $SN/T$.
\begin{figure}[htbp]
\centerline{\includegraphics[width=0.5\textwidth]{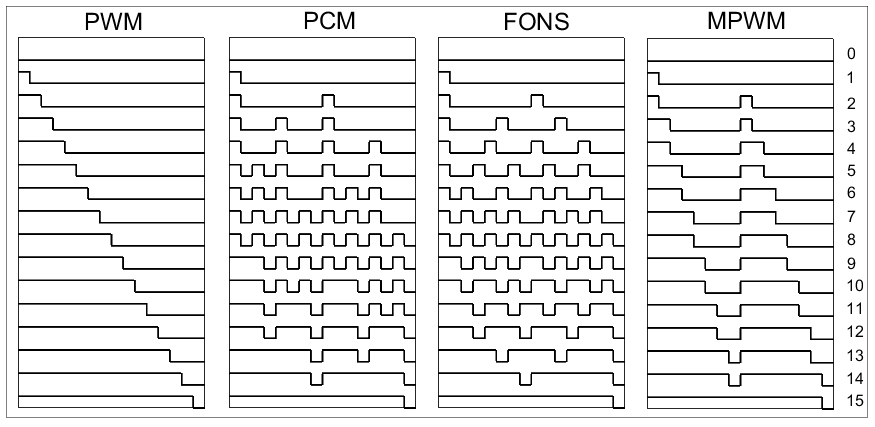}}
\caption{The waveform comparison of the unfiltered digital output.}
\label{wav_compare}
\end{figure}
\begin{figure}[htbp]
\centerline{\includegraphics[width=0.5\textwidth]{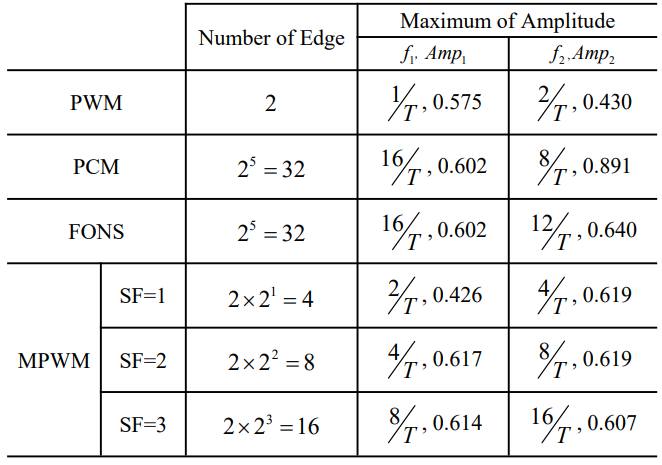}}
\caption{Differences among four waveforms: $f_1$  and $f_2$ are the frequencies with the maximum and the second largest amplitude. Amplitudes have been normalized to their DC value.}
\label{Table1}
\end{figure}
%Fig. \ref{Table1} and Fig. \ref{wav_compare} show the differences between PWM, PCM, FONS, and harmonic components. When compared to the traditional PWM, the low-pass filter used for MPWM has a higher cut-off frequency, which can lower the requirements for filter design in DAC applications. When compared to PCM and FONS, MPWM has fewer edges in the time domain, which can reduce the integral nonlinearity of DAC, as presented in Section III.
Fig. \ref{Table1} and Fig. \ref{wav_compare}  illustrate the distinctions among PWM, PCM, FONS, and harmonic components. In comparison to typical PWM, MPWM utilizes a low-pass filter with a greater cut-off frequency, hence reducing the demands for filter design in DAC applications. MPWM exhibits a lower number of edges in the time domain when compared to PCM and FONS. This characteristic can effectively decrease the integral nonlinearity of the DAC, as discussed in Section III.
\subsection{Performance of MPWM-DACs}
\subsubsection{Static error}
The DC component of the output of an MPWM-DAC can be expressed as in Eq.\ref{u_d}, where $u_d$ is the digital output of MPWM and $D$ is the nominal duty (the actual duty is $D/2^n$, where $n$ is the bit width of the counter).
\begin{equation}\label{u_d}
   \bar{u_d}(D)=\frac{1}{T}\int_{t_1}^{t_1+T}u_d(t,D)dt. 
\end{equation}
The static error of an MPWM-DAC is defined as in Eq.\ref{e_s}, where $u_D$ is the ideal DC voltage value corresponding to the duty D. $U_LSB$ is the voltage corresponding to the DAC’s LSB.
\begin{equation}\label{e_s}
e_s(D)=\frac{\bar{u_d}(D)-u_D}{U_{LSB}}    
\end{equation}
%The main source of the static error for an MPWM-DAC is the power supply error and error caused by edges. Since the “HIGH” value of the output voltage $u_d(D)$ equals the power supply voltage $U_s$, the DC component of MPWM-DAC’s output can have the same error as that of the power supply. In this aspect, MPWM-DAC is identical to the PWM-DAC, PCM-DAC, and FONS-DAC.
The primary cause of static error in an MPWM-DAC is the combination of power supply error and error resulting from edges. Given that the maximum value of the output voltage $u_d(D)$ is equal to the power supply voltage $U_s$, the direct current component of the MPWM-DAC's output can have the same level of inaccuracy as the power supply. MPWM-DAC shares the same characteristics as PWM-DAC, PCM-DAC, and FONS-DAC in this regard.

As for errors caused by edges, since the rising and falling edges of the MPWM output are nonideal, errors can be brought. The nonideal characteristics of edges can be modeled using Trapezoid model \cite{BB1986} as follows:
\begin{equation}\label{e_edge}
 e_{edge}=E(D)(t_{dr}-t_{df})f_{clk}   
\end{equation}
where $E(D)$ is the number of positive (or negative) edges for duty $D$, $t_dr$ is the delay from the half amplitude of the clock signal to that of the MPWM digital output for the rising edge, and  $t_{df}$ is the delay for the falling edge. 

For MPWM, the number of edges is 
\begin{equation}\label{edge}
E_{MPWM}=\left\{\begin{matrix}
D,~~~~~~~~~~~~~~D\leqslant SN~~~~~~~~~~~~
\\ 
SN,~~~~SN<D\leqslant (2^n-SN)
\\ 
2^n-D,~~~~~~D>(2^n-SN).
\end{matrix}\right.    
\end{equation}
For PWM, the maximum number of positive edges is 1, thus has the least edge-caused error. For PCM, and FONS, the maximum number of edges is $2^{n-1}$. When using a counter with 12 bits, there are 2048 edges in most PCM (or FONS) periods, which can cause severe errors. 

For MPWM, the maximum number of edges is SN=$2^{SF}$, where the SF can be chosen from 1 to 10 for a $12$-bit counter, hence MPWM has fewer edges and thus reduces the static error caused by edges proportionally. 

\subsubsection{Integral nonlinearity (INL)}
The integral nonlinearity (INL) of a DAC is the maximum deviation of the actual analog output from the ideal output. The INL for low-cost DAC can be calculated as in \cite{Zander1985}: 
\begin{equation}\label{e_INL_edge}
e_{INL}=\max_{D}\mid e_{edge}(D))\mid
\end{equation}
For MPWM-DAC, the INL is as follows: 
\begin{equation}\label{e_INL_MPWM}
    e_{INL(MPWM)}=\mid 2^{n-1}(t_dr-t_df)f_{clk}\mid,~SF=1,...,n-2.
\end{equation}
For PWM-DAC, PCM-DAC, and FONS-DAC, the INL is as follows:
\begin{equation}\label{eINL}
e_{INL}=\left\{\begin{matrix}
\mid (t_dr-t_df)f_{clk}\mid,~PWM~~~~~~~~~~~~~~~~~~~~~~~~~~ 
\\ 
\mid 2^{n-1}(t_dr-t_df)f_{clk}\mid,~PCM~\&~FONS.
\end{matrix}\right.    
\end{equation}
Since $SF$ ranges from $1$ to $n-1$, MPWM can have much better INL performance than PCM-DAC and FONS-DAC. 
\subsubsection{Differential nonlinearity (DNL)}
The Differential nonlinearity(DNL) for DAC is defined as the maximum voltage deviation of the DAC output between two adjacent digital inputs in terms of an ideal output voltage step corresponding to 1 LSB. It is calculated as:
\begin{equation}\label{eDNL}
    e_{DNL}=\frac{\bar{u_d}(D+1)-\bar{u_d}(D)}{U_{LSB}}-1.
\end{equation}
Substitude Eq.\ref{e_edge} and Eq.\ref{edge}, the DNL for MPWM is 
\begin{equation}\label{eDNL_MPWM}
    e_{DNL(MPWM)}=\mid (t_dr-t_df)f_{clk}\mid .
\end{equation}
The DNL performance of MPWM is the same as that of PWM-DAC, PCM-DAC, and FONS-DAC. To improve the DNL of MPWM, the output stage needs to be as symmetric as possible, i.e. having the same characteristics for the rising and falling edges, to reduce the $\mid t_dr-t_df\mid$ item. This symmetry is crucial when the clock frequency is high. Due to its operating principle, MPWM-DAC is monotonic.
\subsubsection{Dynamic Characteristics}
\begin{figure}[htbp]
\centerline{\includegraphics[width=0.5\textwidth]{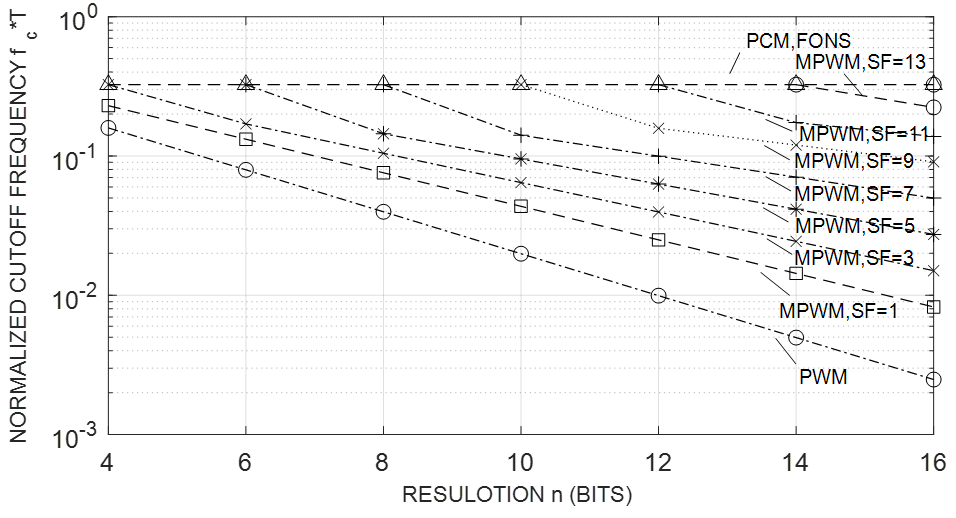}}
\caption{Normalized cutoff frequency vs. DAC resolution, using second order Butterworth low pass filter.}
\label{Normalized_cutoff}
\end{figure}
For dynamic characteristics analysis, getting a general analytical computation for the output from MPWM-DAC is quite difficult, therefore, we use numerical computation. The ripple of the output of an MPWM-DAC with second order Butterworth low pass filter is calculated to get normalized cutoff frequency $f_c T$ of the filter for different $SF$ in Fig. \ref{Normalized_cutoff}.
The $f_cT$ increases with the increase of SF. For instance, when designing a 12-bit DAC, $f_c T$ is 0.01 for PWM-DAC, and for MPWM-DAC it increases to 0.04 and 0.1 when $SF$ equals 3 and 7 respectively. Thus the cutoff frequency is 4 and 10 times that of PWM-DAC, and the settling time is 1/4 and 1/10 of that of PWM-DAC. 
\section{FPGA DESIGN AND RISC-V SOC}
\begin{figure}[htbp]
\centerline{\includegraphics[width=0.5\textwidth]{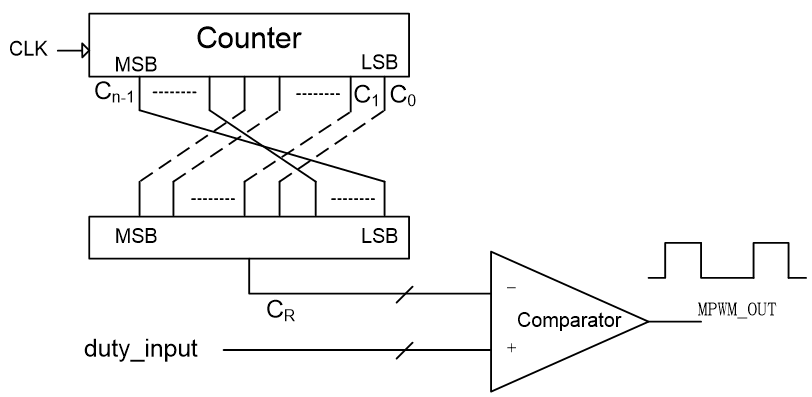}}
\caption{MPWM Circuit.}
\label{MPWM_circuit}
\end{figure}
\subsection{MPWM circuit design}
The block diagram for an MPWM circuit is shown in Fig.\ref{MPWM_circuit}, where the bit width of the counter is $n$. For an MPWM wave with $SF$, the output of the counter is rearranged according to
$ C_R$=$\{C_{n-SF-1},C_{n-SF-2},\cdots,C_0,C_{n-SF},\cdots,C_{n-2},C_{n-1}\}$, where $C_0$ to $C_{n-1}$ are the output of the counter, and $C_R$ is the result of rearrangement. The duty target is input from $duty_input$, which ranges from 0 to $2^{n-1}-1$. Then $duty_input$ is compared with  $C_R$, and MPWM outputs according to Eq. \ref{MPWM}.
\begin{equation}\label{MPWM}
MPWM\_OUT=\left\{\begin{matrix}
1,~duty\_input \geq C_r
\\ 
~0,~duty\_input < C_r.
\end{matrix}\right.      
\end{equation}
\subsection{HR-MPWM design in FPGA}
\begin{figure}[htbp]
\centerline{\includegraphics[width=0.5\textwidth]{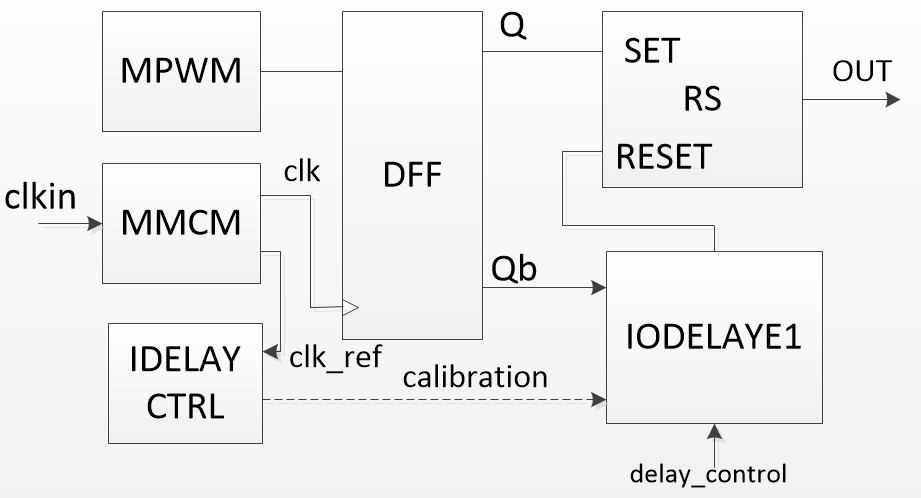}}
\caption{Block diagram of HR-MPWM in FPGA. }
\label{HR-MPWM}
\end{figure}
To increase the DAC’s resolution, one can increase the bit width of the counter, but the clock rate needs to be raised simultaneously, or the DAC’s speed will decrease proportionally. In many cases, it is very difficult to raise the clock rate due to the process limit or power consumption. To solve this problem, the circuit to generate a pulse with a resolution better than one clock cycle can be designed. An MPWM with this kind of circuit is named high-resolution MPWM, i.e. HR-MPWM. When implemented in Xilinx FPGA, the HR-MPWM circuit is shown in Fig. \ref{HR-MPWM}, where $IODELAYE1$ is the $I/O$ delay element and can allow a fine delay-time $t_{d}=1/(32\times2\times f_{clk_{ref} })$. Thus, this circuit can provide an additional $6$-bit resolution for MPWM. Considering the HR-MPWM in the final SoC needs only $4$ additional bits, the design in FPGA uses a delay tap out of every $4$ taps. 

\subsection{SoC Structure with the proposed DAC}

\begin{figure*}[htbp]
\centerline{\includegraphics[width=0.8\textwidth]{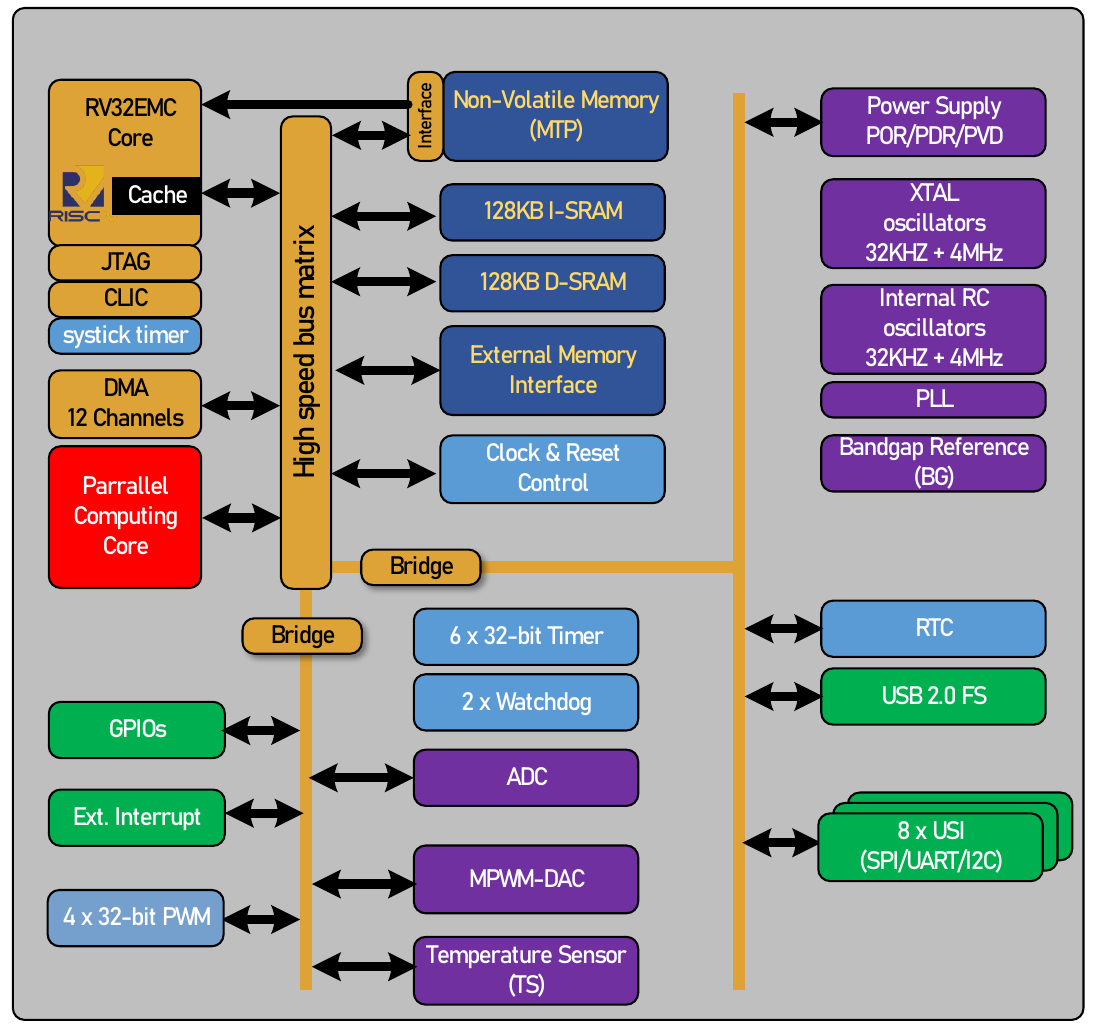}}
\caption{A brief design architecture of the RISC-V SoC for THz IoT devices.}
\label{RISCBlock}
\end{figure*}

The MPWM-DAC is integrated into a RISC-V SoC for IoT applications. Its position within the SoC level range is shown in Fig. \ref{RISCBlock}, where it is located alongside its associated auxiliary modules within the "MPWM-DAC".

Control over the entire system is exerted by an open-source 32-bit RISC-V core. For the support of complex vector operations, commonly required in AI acceleration, a specialized parallel computing core is strategically deployed. Configuration of the proposed MPWM-DAC module is undertaken by the RISC-V core, facilitated through a high-speed bus linking them. The responsibility for managing massive data transportation falls to DMAs. Wireless communication capabilities are introduced via an off-chip Wi-Fi chip, which is connected to a UART port. The system's memory requirements are addressed with on-chip 128KB instruction SRAM and an equal capacity for data SRAM. Additionally, the External Memory Interface is employed to access further memory, with the capacity to address nearly the entirety of the remaining 4GB.

\subsection{Test in FPGA}
\begin{figure}[htbp]
\centerline{\includegraphics[width=0.5\textwidth]{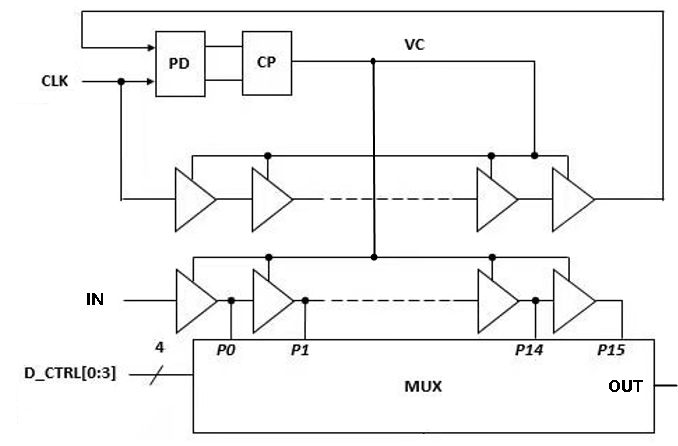}}
\caption{DLL design in the RISC-V SoC.}
\label{DLL}
\end{figure}
\begin{figure}[htbp]
\centerline{\includegraphics[width=0.5\textwidth]{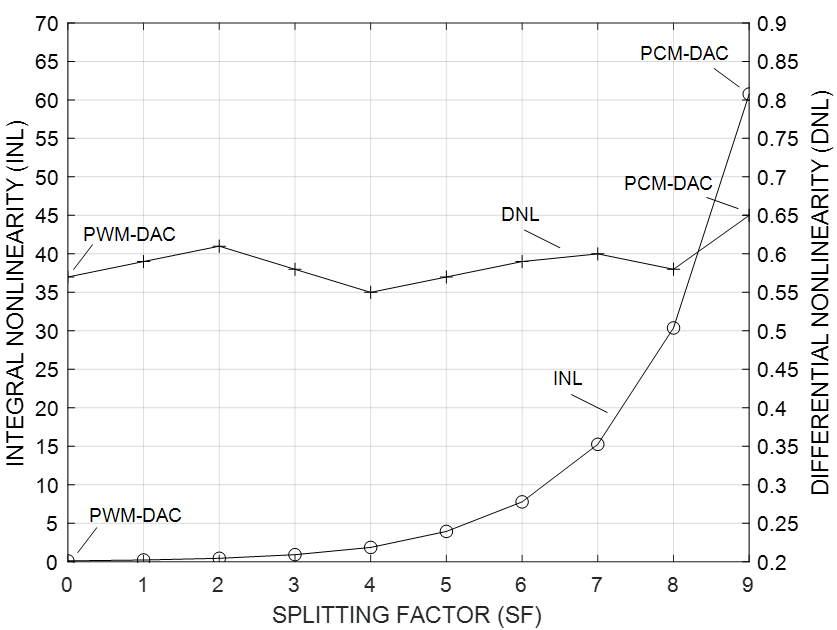}}
\caption{INL and DNL performance for MPWM.}
\label{INL_DNL}
\end{figure}

The prototyping SoC with MPWM-DAC is implemented in an Xilinx ARTIX-7 FPGA (XC7A200TFBG484-3), as shown in Fig. \ref{FPGA}. The second-order Butterworth filter is connected to the output of the DAC. The INL and DNL for an MPWM-DAC with $10$-bit resolution are measured, and compared with PWM-DAC and PCM-DAC as presented in Fig. \ref{INL_DNL}.

\begin{figure}[htbp]
\centerline{\includegraphics[width=0.5\textwidth]{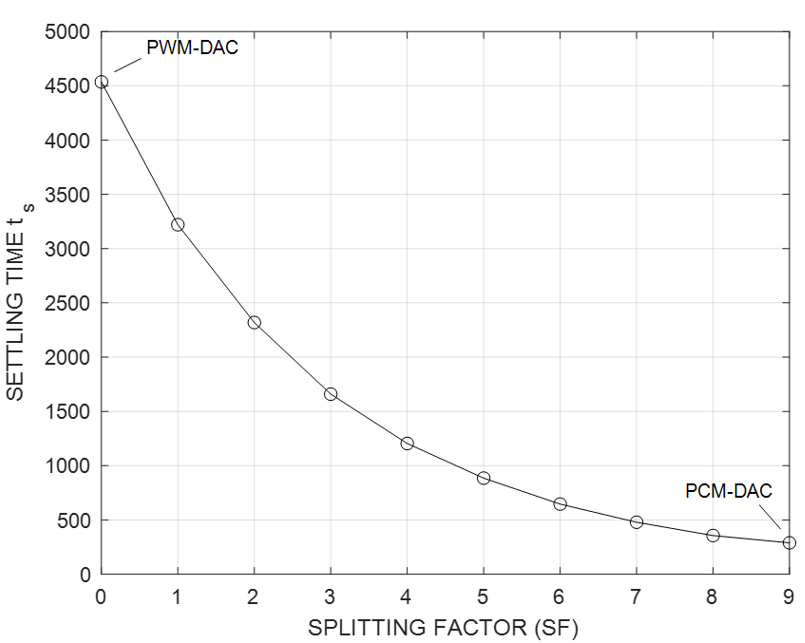}}
\caption{Setting time performance for MPWM.}
\label{Setting_time}
\end{figure}

These results show that MPWM-DAC has improved the INL performance greatly when compared to PCM-DAC, and has a shorter settling time than PWM-DAC in Fig. \ref{Setting_time}. 
When designing an MPWM-DAC, one can get a higher accurate DAC by choosing a smaller SF or get a faster DAC by selecting a larger SF.   
\subsection{RISC-V SOC using MPWM-DAC for THz IoT devices}
\begin{figure}[htbp]
\centerline{\includegraphics[width=0.5\textwidth]{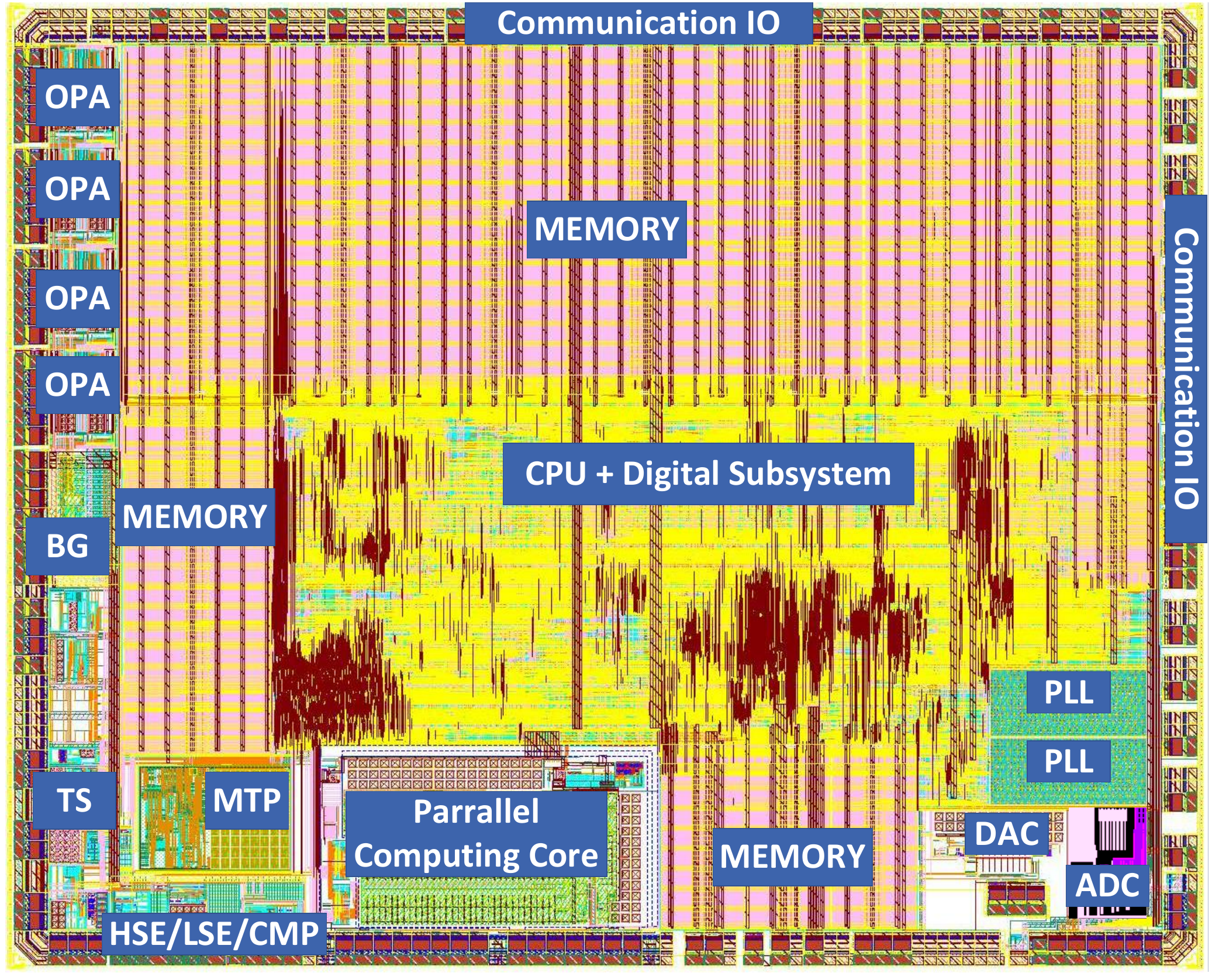}}
\caption{Layout of the RISC-V SoC.}
\label{RISC}
\end{figure}

In the SoC, the HRMPWM module is a fully customized design, which comprises a typical PWM generator and delay elements calibrated by a DLL. The DLL is applied to get 16 fine phases, refer to Fig. \ref{DLL}. After the DLL is locked, $vc$ is generated and applied to the signal path for MPWM, which has the same delay element as in DLL. A multiplexer circuit with the same delay in every branch is employed to get the required phase. 
\begin{figure}[htbp]
\centerline{\includegraphics[width=0.5\textwidth]{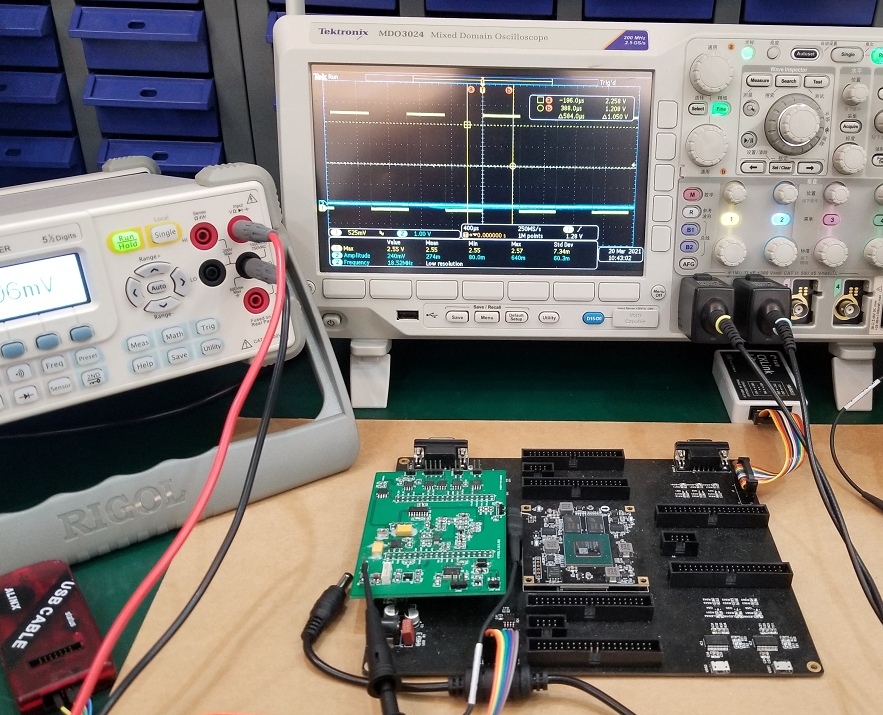}}
\caption{Experimental setup with FPGA.}
\label{FPGA}
\end{figure}
The laboratory experiment for the FPGA test is illustrated in Fig. \ref{FPGA} and the layout of the MPWM SoC is shown in Fig. \ref{RISC}. It is fabricated (taped out in March 2021) in a 180$nm$ CMOS process with a size of $3.8mm\times4.7mm$. The size of HRMPWM is $0.53mm\times0.27mm$. 
\section{CONCLUSION}
THz communications are expected to have a crucial impact on the development of wireless systems in the sixth generation (6G). This study provides a comprehensive analysis of THz IoT devices. In addition, the MPWM modulation mechanism, introduced in an MPWM circuit with a $n$-bit counter, may exhibit $(n-1)$ distinct configurations and waveforms. The MPWM-DAC is developed using the MPWM technology, offering superior conversion speed compared to PWM-DAC, and higher accuracy than the PCM-DAC. By virtue of its versatility, the designer is able to modify their design to suit unique applications, resulting in a superior outcome. An implementation using a delay line has been developed in order to enhance the resolution of MPWM-DAC. The MPWM-DAC is included in a RISC-V SoC and evaluated on an FPGA. The SoC is specifically engineered using a 180$nm$ CMOS fabrication technique.

\bibliographystyle{IEEEtran}
\bibliography{main}

%\EOD

\end{document}